\documentclass[aps,prl,twocolumn,superscriptaddress,showpacs]{revtex4-1}
\usepackage{graphicx}
\usepackage{amsmath}
\begin{document}


\title{Homometry in the light of coherent beams}

\author{Sylvain Ravy}
\email[]{ravy@synchrotron-soleil.fr}
\affiliation{Synchrotron-Soleil, L'Orme des merisiers, Saint-Aubin BP48, 
91192 Gif-sur-Yvette, France}
\homepage{http://www.synchrotron-soleil.fr/Recherche/LignesLumiere /CRISTAL}

\date{\today}

\begin{abstract}
Two systems are homometric if they are indistinguishable by diffraction.
We first make a distinction between Bragg and diffuse scattering homometry, 
and show that in the last case, coherent diffraction can allow the diffraction
diagrams to be differentiated.
The study of the Rudin-Shapiro sequence, homometric to random sequences, allows
one to manipulate independently two-point and four-point correlation functions,
and to show their effect on the statistics of speckle patterns.
Consequences for the study of real materials are discussed.

\end{abstract}


\maketitle

\section{Introduction}

The possibility to shape coherent X-ray beams from synchrotron light 
sources \cite{Sutton} and to get naturally coherent beams from X-ray free 
electron lasers (XFEL) \cite{Vartanyants2011}, has revolutionized the way 
X-ray diffraction experiments are performed and analyzed.
One of the most fascinating property of coherent diffraction is the possibility
to measure speckle patterns \cite{Livet}, which are much more informative
than the diffuse scattering obtained by classical diffraction. 
Together with the development of novel sources, phase retrieval algorithms have
also emerged, allowing the reconstruction of the diffracting objects under 
certain conditions \cite{Miao,Rodenburg}.
However, the reconstruction of a structure is not always possible nor necessary
to study the physics of materials.
For example, measuring correlation lengths close to phase 
transitions \cite{Ravy2007} or slow dynamics with 
X-ray photon correlation spectroscopy \cite{Livet,Grubel2004} does not require
the full reconstruction of the system under study.

The purpose of this letter is to show that statistical analysis of speckle 
patterns can yield information on orders hidden to conventional X-ray 
analysis.
In this respect, we are in line with recent works showing that
four-point intensity cross-correlation of speckle patterns 
can uncover "hidden symmetries" present in colloidal glasses 
\cite{Wochner2009,Altarelli2010} or magnetic systems 
\cite{Su2011}.

Our approach uses the concept of homometry, {\it i.e.} the property of 
different systems to exhibit same diffraction patterns.
Separating out the scattered intensity expression into three terms allows 
one to show that homometry can occur at different levels.
We then put the emphasis on diffuse scattering homometry, that we discuss with the 
help of the well-known Rudin-Shapiro sequence \cite{Axel1992,Baake2009}.

\section{Coherent diffraction}

Let us first give a general expression of the intensity scattered at scattering 
vector $q$, by a 1D periodic $N$-site lattice decorated by two atoms 
$A$ and $B$, of scattering factor $f_A$ and $f_B$, in proportion $x$ and
$1-x$ respectively.
Generalization to 2D, 3D, multi-atomic basis, displacement disorder, or disorder 
of the second kind \cite{Guinier94} is straighforward.
Following ref. \cite{Guinier94}, the diffracted intensity is given by the formulae:
\begin{equation}
I(q)=\sum_{n,n'}f_nf_{n'}e^{iq(n'-n)}=\sum_{m}\sum_{n}f_nf_{n+m}e^{iqm}.
\label{intensity}
\end{equation}
The \emph{ensemble} average of the cross-product 
$\langle f_0f_m \rangle$ is then introduced:
\begin{equation}
\frac{1}{N_m}\sum_{n}f_nf_{n+m}=\langle f_0f_m \rangle+\Delta_m,
\end{equation}
where $N_m$ is the $m$-dependant number of terms of the sum $\sum_{n}$.
The $\Delta_m$ term, usually negleted in textbooks, is due to finite-size 
fluctuations of the spatial average with respect to the ensemble one.

Further introduction of $\Delta f_m=f_m-\langle f \rangle$ allows one to get the 
three components of kinematic diffraction:
\begin{subequations}
\begin{eqnarray}
I_B(q)&=& \langle f \rangle^2 \sum_{m} N_m e^{iqm}
\label{Ia}\\
I_{DD}(q)&=& \sum_{m} N_m\langle \Delta f_0\Delta f_m \rangle e^{iqm}
\label{Ib}\\
I_S(q)&=& \sum_{m} N_m\Delta_m e^{iqm}
\label{Ic}
\end{eqnarray}
\label{Iabc}
\end{subequations}

The first term gives the intensity of the Bragg reflections, and the fringes
due to finite size effects.
For a crystal of $N$ cells of structure factor $F(q)$, it can be written as:
\begin{equation}
I_B(q) = |\langle F(q) \rangle|^2 \frac{\sin^2q(N+1)/2}{\sin^2q/2}
\label{fringes}
\end{equation}
In practice, the fringes given by the sine functions are only visible with a 
coherent beam illumination (for conditions of observation and examples 
see \cite{Livet}).

The second term is the diffuse scattering intensity,
which only depends on pair correlation function (CF). 
For \emph{random} disorder, it reduces to the well-known Laue formula:
\begin{equation}
I(q)=N x(1-x)(f_A-f_B)^2,
\label{Laue}
\end{equation}
where \emph{random} refers to the vanishing of the pair CF 
({\it i.e.} $\langle \Delta f_0\Delta f_m \rangle = 
\langle \Delta f_0\rangle\langle\Delta f_m \rangle=0$ for $m\ne0$.)

The third term gives rise to speckles.
Like fringes, speckles only exist if the incident beam 
is coherent enough, \emph{and} if the system does not explore too many 
configurations during acquisition time $T$ (non-ergodicity condition 
$\langle \Delta_m \rangle_T \ne 0$).
Interestingly enough, working out of coherent conditions has the effect of 
averaging out $\Delta_m$, which yields \emph{ensemble}
averaged quantities. 

\begin{figure}
\includegraphics[width=6.0cm]{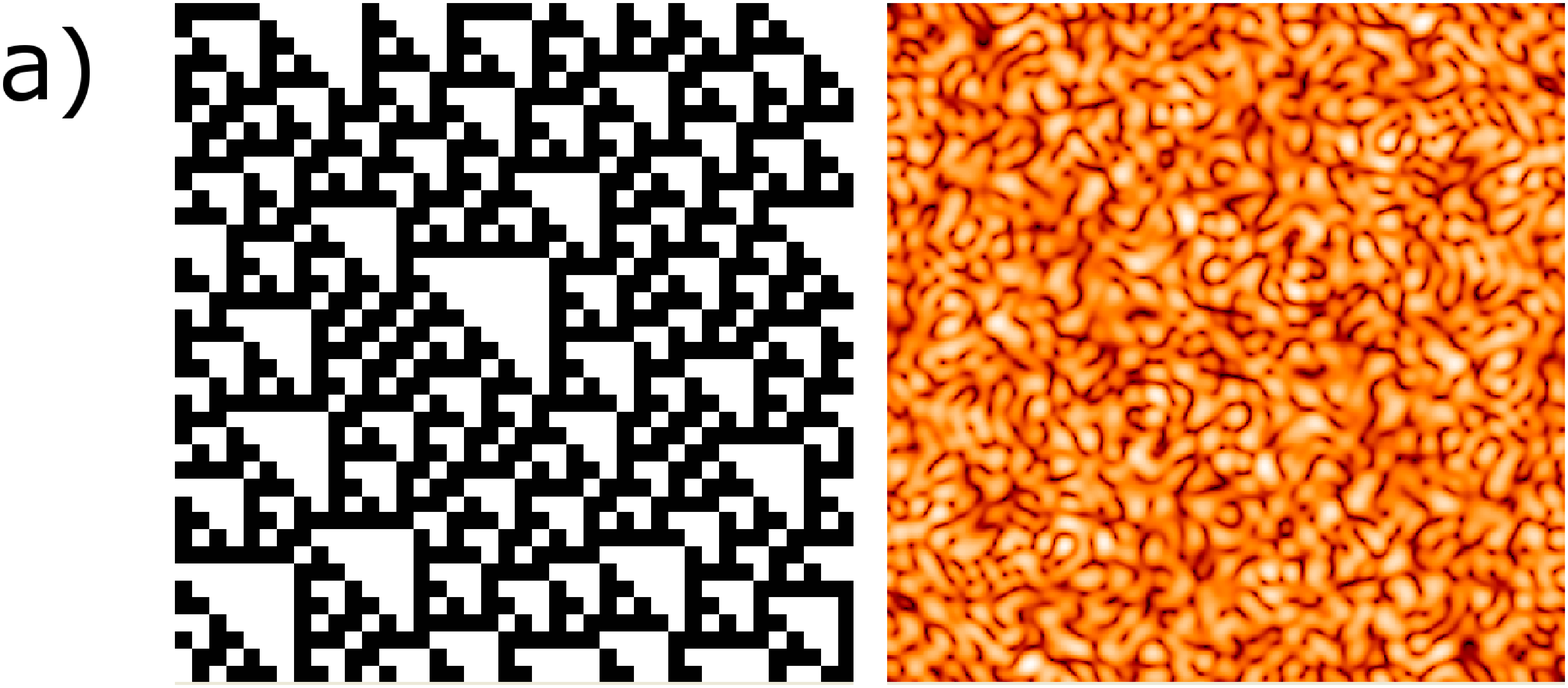}
\vskip 0.1cm
\includegraphics[width=6.0cm]{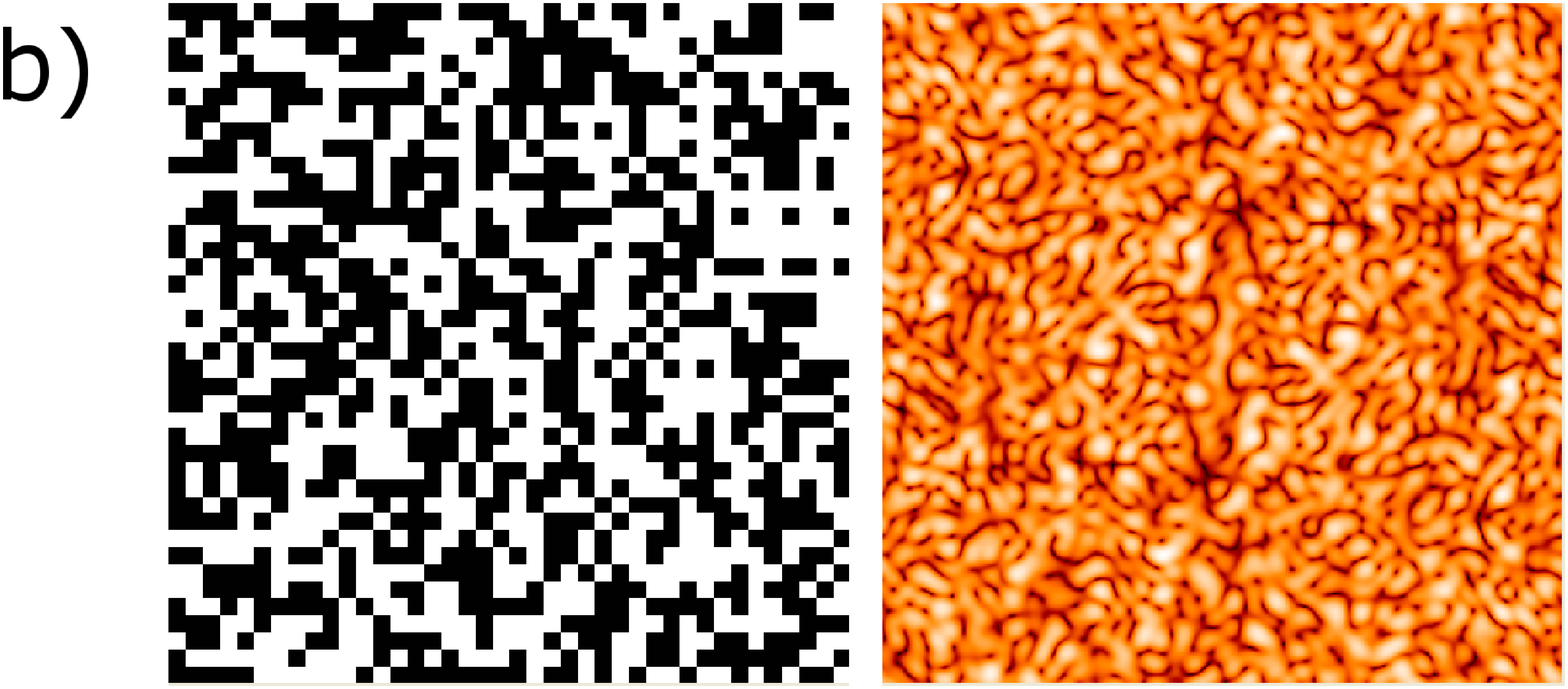}
\caption{\label{W}(left) $(40\times40)$ lattices of +1 and -1 in equal proportion 
and (right) diffraction patterns in the first Brillouin 
zone (log scale) for a) a triplet SRO lattice \cite{Welberry1994} b) a random
lattice.
Reciprocal nodes are in the corners.}
\end{figure}

\section{homometry}
Homometry - etymologically \emph{same distance} - is a word coined by A. 
Patterson \cite{Patterson,Patterson44}, to describe the property of 
non-congruent sets of points to possess the same pair distances 
(or the same difference sets) \cite{Senechal}. 
Homometric sets have thus the same diffraction pattern, as Eq.~(\ref{intensity}) 
demonstrates.
A simple example of homometry is given by the two sets $S=\{0,1,4,10,12,17\}$ and
$S'=\{0,1,8,11,13,17\}$ \cite{Senechal}.
Indeed, the structure factors $F(q)$ of the two sets have the 
same magnitude for \emph{all} q-vectors, but not the same phase.
Hence, the lost of the phase makes these sets \emph{indistinguisable} by X-ray 
diffraction.

However, because solid state physics deals with materials, 
the above definition turns out to be too restrictive.
Eqs (\ref{Iabc}) allows to distinguish between Bragg (B) homometry, diffuse 
scattering (D) homometry and coherent diffraction (C) homometry. 

\section{Bragg homometry}

B-homometry decribes crystals with different basis but same Bragg 
intensities \cite{Patterson,Patterson44}.
To illustrate that, let us consider the examples of 1D homometric 
crystals presented in \cite{Patterson44}, of unit cells size equal to 8 
and atomic positions given by $H=\{0,3,4,5\}$ and $H'=\{0,4,5,7\}$.
Structure factors, readily calculated as:
\begin{eqnarray}
F_H(q)&=& 1+2\cos q+\cos 4q\\
F_{H'}(q)&=& 2(\cos2q+\cos4q),
\end{eqnarray}
have the same amplitude squared \emph{at} the Bragg positions $q~=~2\pi h/8$ 
but not \emph{out of} Bragg positions.
Eq (\ref{fringes}) shows that the fringes intensity, revealed by coherent 
diffraction, gives out-of-Bragg values of $|F(q)|^2$ which, at least in theory, 
allows to distinguish $H$ and $H'$, and solves the B-homometry issue.
This is well-known and corresponds to the oversampling 
requirement of the phase retrieval algorithms \cite{Sayre,vanderVeen}. 
It is clear however that if the atomic basis is homometric itself, 
like {\it e.g.} in crystals with $S$ or $S'$ basis, the problem cannot be solved, 
coherence or not.

\begin{figure}
\includegraphics[width=7.25cm]{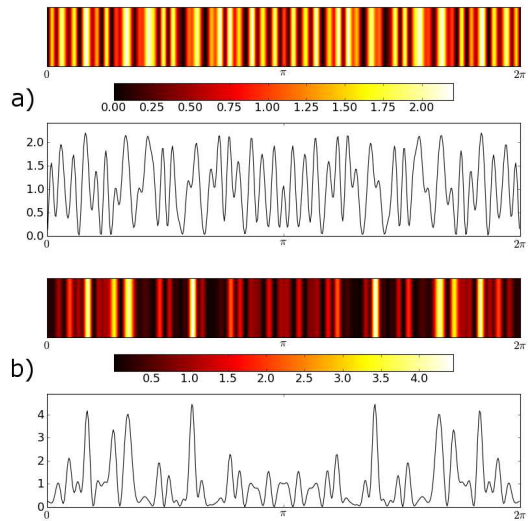}
\caption{\label{RS_1}Diffraction patterns and associated intensity variations 
from ($(N,M)=(64,512)$) a) RS and b) BS sequences.
The vertical broadening of the patterns are for visual convenience.}
\end{figure}

\section{Diffuse scattering homometry}
Surprising examples of D-homometry were designed by Welberry {\it et al.}
\cite{Welberry1977,Welberry1994}.
They consist in substitutionnally disordered lattices with 
triplet (or quadruplet) short range ordered (SRO) CF \footnote{In the following, 
we consider that an order parameter is short-range ordered (SRO) if its 
associated CF $g(n)$ vanishes at infinity and long-range 
ordered (LRO) otherwise.}, but
zero two-point correlations (Fig. \ref{RS_1}a).
Diffraction diagram of these lattices present the same Bragg and diffuse 
scattering intensity \cite{Welberry1994}. 
But, as shown in Fig. \ref{RS_1}, their speckle patterns are different, 
which breaks the D-homometry.

A more tractable (though subtle) example of D-homometry is provided by the 
geometrically ordered (GO) \cite{Gratias2005} Rudin-Shapiro (RS) sequence 
\cite{beyond95}, whose generic term $\sigma_n$ can be written \cite{Baake2009}:
\begin{equation}
\sigma_{4n+l} = \left\{ 
\begin{array}{lcl}
\sigma_{n} & \mbox{for} & l=0,1 \\ 
(-1)^{n+l}\sigma_{n} & \mbox{for} & l=2,3 
\end{array}\right.
\text{ with }\sigma_{0} = 1.
\label{defRS}
\end{equation}

This sequence has become famous \cite{Axel1992,Hoffe,beyond95,Gratias2005} 
because, thought GO, it is D-homometric to randomly distributed sequences 
(sometimes called Bernoulli sequences (BS) \cite{Hoffe}), with the same
$4Nx(1-x)$ diffuse scattering intensity (Eq. \ref{Laue}).
In other words, its two-point CF
$g_2(n)=\overline{\sigma_0\sigma_n}$ is zero for $n\ne0$ (\emph{spatial} average).

\begin{figure}
\includegraphics[width=7.25cm]{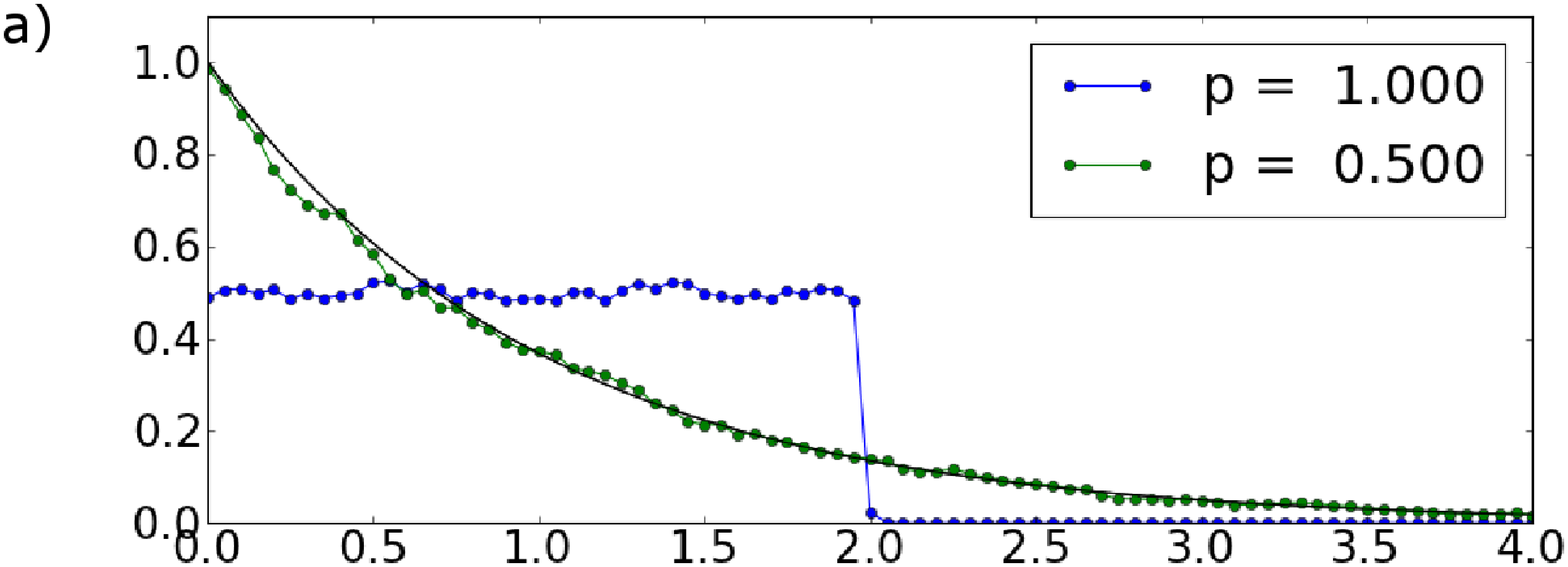}
\includegraphics[width=7.25cm]{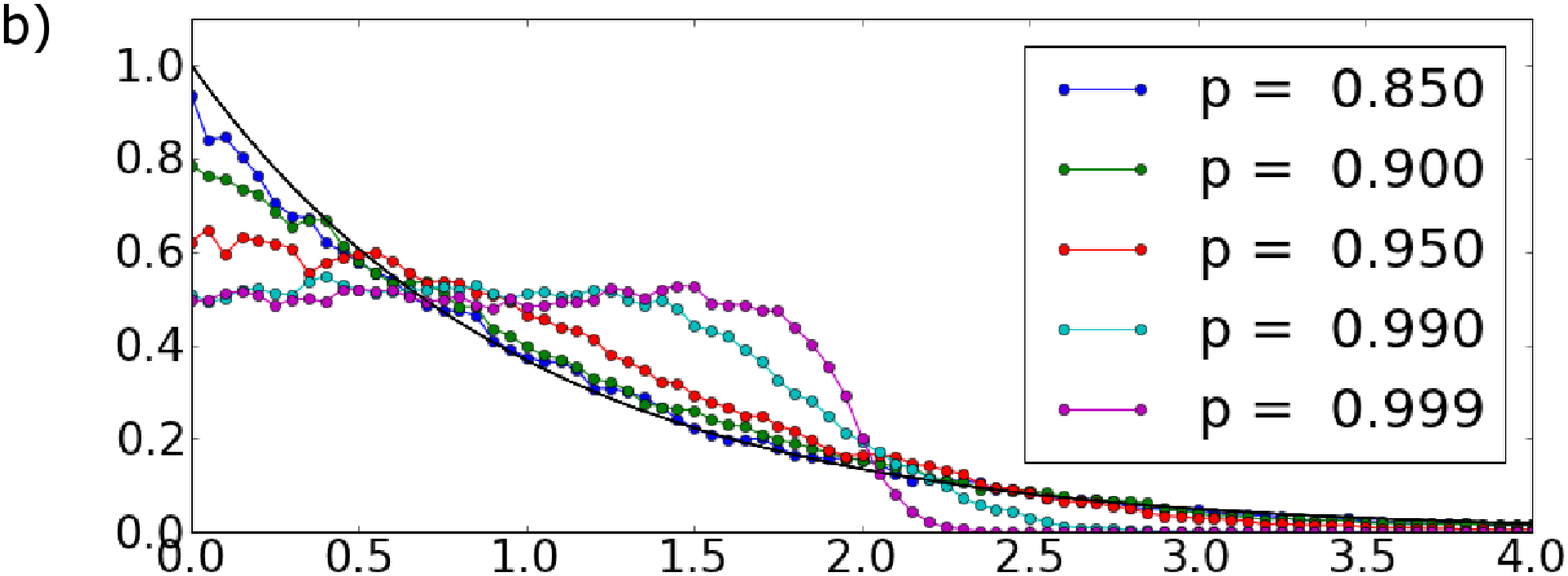}
\includegraphics[width=7.25cm]{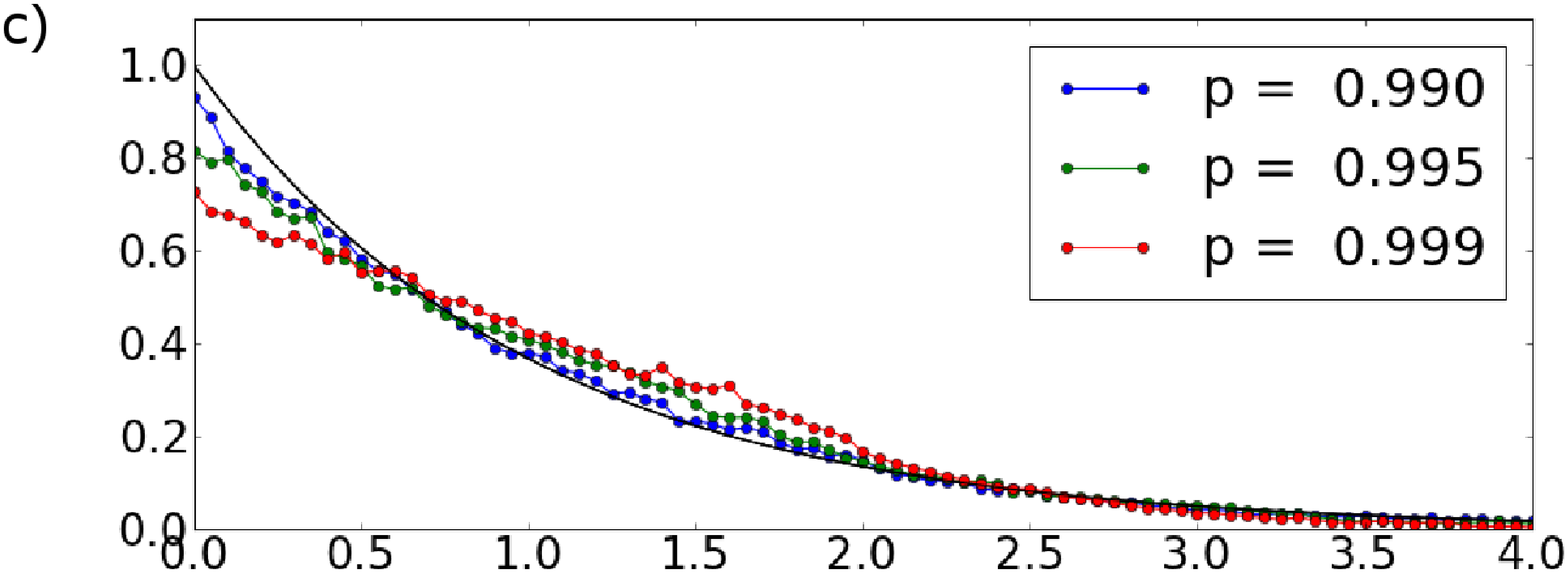}
\caption{\label{DP}Probability densities $P(I)$ of diffraction patterns 
of $(N,M)=(4096,4096\times128)$: a) BS (green) and RS sequence (blue),
b) sequences obtained by Bernoullization c) $g_4$-SRO sequences for 
$\xi \sim$ N/37 ($p=0.99$), N/13 ($p=0.95$) and N/3 ($p=0.999$).
The line indicates the random negative exponential law.}
\end{figure}

Let us now consider the coherent diffraction of RS and BS sequences.
In what follows, we present Fast Fourier Transforms (FFT) computations 
of sequences of length $N$, zero-padded up to a value $M\gg N$ to 
clearly see the speckles.
For the RS sequence, because $x\ne0.5$ and depends on its length $N$, 
we have always susbstracted the average value $2x-1$ to all 
the terms in order to get rid of the Bragg intensites.
BS of 1 and -1 in equal proportion were computed with 
the python pseudo-random number generator.
The squared value of the FFT $I(q)$ are normalized by 
$4Nx(1-x)$ in order to get $\overline{I(q)}=1$.

Figure \ref{RS_1} shows the diffraction patterns of RS and 
BS sequences in the first Brillouin zone ($0<q<2\pi$).
Inspection of these patterns shows that, although their average value
is the same, the speckles repartition is remarkably different.
In particular, it is clear that the BS pattern exhibits much more spikes, while
the RS pattern is more homogenous and regular 
(no low-intensity speckles more regularly spaced).
To quantify this observation, we studied the statistics of both speckle patterns 
by calculating their probability density of intensity $P(I)$.
It is known that for a random media (see {\it e.g.} \cite{Dainty1976}), 
the intensity distribution has a negative exponential distribution given by:
\begin{equation}
P(I)=\frac{1}{\overline{I}}\exp{-\frac{I}{\overline{I}}},
\end{equation}
which in our case reduces to $P(I)=\exp(-I)$.

\begin{figure}
\includegraphics[width=7.25cm]{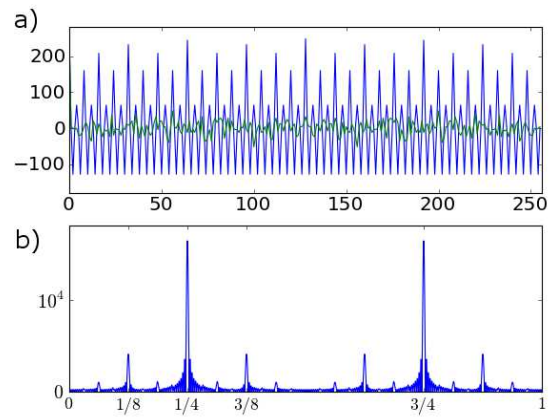}
\caption{\label{RS_4}a) Computed $g_4(n)$ using periodic boundary conditions 
for $N=256$ RS sequence (blue) and a BS sequence (green).
b) Magnitude of RS $\hat{g}_4(q)$ as a function of $h=q/2\pi$.
}
\end{figure}

Figure \ref{DP}a) shows the probability densities $P(I)_{RS}$ and 
$P(I)_{BS}$ for the RS and BS sequences.
While $P(I)_{BS}$ follows quite well the negative exponential law, as 
expected, it is not the case for $P(I)_{RS}$.
Though the precision of $P(I)_{RS}$ depends on $N$, it is well 
approximated by the step function $P(I<2)=0.5$.
This statistics, which means that intensities lower than 2 occur with the same 
probability, explains the homogeneous aspect of the diffraction pattern.
The reason for this unusual statistics is not clear, but
undoubtedly comes from the GO nature of the RS sequence.
The presence of order, invisible through diffuse scattering, is revealed 
by the statistics of the speckle pattern, breaking the D-homometry in a 
quantifiable way.

In order to test the robustness of the $P_{RS}(I)$ behavior with respect 
to disorder, we first quantify the degree of order of the RS by one 
of its quadruplet CF:
\begin{equation}
g_4(n)=\overline{\sigma_0\sigma_1\sigma_n\sigma_{n+1}}.
\end{equation}
Indeed, we found numerically that at variance with the BS, $g_4(n)$ is LRO for 
the RS sequence (Fig. \ref{RS_4}a)\footnote{We checked that RS triplet CF
$g_3(n,n')=\overline{\sigma_0\sigma_n\sigma_{n'}} \sim 0$ and do not show any 
structure, so that quadruplet are the first relevant RS high-order CF.
Amongst them, we chose $g_4(n)$ because it is reminiscent of the
very definition of $\sigma_n$.}.
This is confirmed by the behavior of its FFT $\hat{g}_4(q)$ (Figure \ref{RS_4}b), 
which exhibits well defined peaks indexed by the basis vectors 
$\{\frac{h_i}{4.2^i} | h_i \in \{-1,1\},i\in N\}$ \footnote{This behavior has been 
checked up to the $N=2^{14}$ RS sequences} characteristic of 
limit-periodic functions \cite{Baake2010}.
By analogy with two-point orders, we define $\eta_4\equiv \hat{g}_4(\pi/2)/N^2=1/4$ 
as the order parameter of this sequence.

We first decreased $\eta_4$ while keeping $g_4$-LRO
\emph{and} $g_2$-disorder by using the "Bernoullization" procedure as defined 
in \cite{Baake2009}.
It consists in changing the sign of each $\sigma_n$ with probability $p$, 
in order to build sequences intermediate between
the pure RS ($p=0,1$) and BS ($p=0.5$) sequences.
The order parameter $\eta_4$  was found to vary as 
$\eta_4(p)\simeq (1-2p)^4/4$. 

Typical density probabilities density shown in Fig. \ref{DP}b), 
exhibit a continuous evolution as a function of $p$.
The step-function behavior is rapidly lost as $p \to 0.5$, with the
best sensitivity close to the small intensity values $P(I=0)\equiv P_0$.
Simulations show that $P_0(p)$ closely follows $1-2\eta_4(p)$, 
and that sizeable deviation from the normal law starts from $p\gtrsim 0.8$.

Effect of $g_4$-SRO on the speckle pattern was studied 
by shifting the sequence by two lattice periods 
at certain points, randomly selected with probability $p$. 
This ensures to achieve $g_4$-SRO, clearly seen by the broadening of 
the $\hat{g}_4(\pi/2)$ peak, \emph{while keeping} the $g_2(n)$ correlation 
to zero.
Figure \ref{DP}c) shows the results for different $p$ values, correponding to
the average distances between faults $\xi$ given in the caption.
We checked that $P(I)_{RS}$ is unaffected for $\xi \lesssim N/40$. 
This shows that the extent of the four-point order has also an effect on the 
density probability, much larger than in the Bernoullization 
procedure.

Finally, let us mention that the problem of C-homometry, which is somehow 
the \emph{true} homometry, is clearly related to the unicity of inverse 
problems, which is beyond the scope of this paper.
It is important to note, however, that the use of ptychography \cite{Rodenburg}, 
in which diffraction patterns are obtained by shifting illumination on the sample, 
can solve difficult problems of phase retrieval \cite{Guizar}, 
including C-homometry.

\section{Discussion}

It might seem pointless to discuss the problem of homometry while phase retrieval
algorithms and ptychography can provide the full structural information, 
including high-order correlation functions.
However, the full measurement of 3D speckle patterns is time consuming and 
in many situations it is not possible to get the data needed for such inversions.
Methods of speckle analyses on quickly measured diagrams are thus needed to get 
novel information on the materials.

As stressed in ref. \cite{Schneider}, real cases of true B-homometry are rare.
On the contrary, because disorder is concerned, D-homometry is very frequent 
especially when systems are large. 
Moreover, it is related to high-order CF, which are hardly 
accessible to experiments \cite{Wochner2009,Su2011,Treacy}\footnote{Let us 
emphasize that the techniques developed in
ref. \cite{Treacy} to get high-order CF require sample scanning, as in 
ptychography.}.
However, although D-homometric systems have almost always different 
speckle patterns, it is not yet clear whether this difference is quantifiable.
Indeed, though the probablity density 
$P_{RS}(I)$ of our test bed sequence is strikingly different from its 
homometric BS, the observed effects are very sensitive to disorder, 
and are almost invisible when four-point correlation lengths are too short.
In this respect, we checked that the probablity density of the 
lattice shown in Fig. \ref{W} does not present sizeable deviation from the 
normal law.
Qualitatively, the "spikiness" of speckle patterns is quickly reinforced
by the introduction of randomness, which makes deviation to the negative 
exponential curve delicate to observe.

Consequently, though it is tempting to conclude from this study that high-order 
correlation functions are directly observable through speckle statistics analysis, 
much theoretical work is still needed to find the \emph{relevant}
parameters controlling the statistics.
Such an effort could be supported by more sophisticated analyses such as 
the use of second-order (or higher) probability functions \cite{Dainty1976}.
The simple fact that the lattices of Fig. \ref{W} can be reconstructed with 
minimum information shows that high-order CF are somehow hidden in the speckle 
repartition.

An experimental difficulty lies in the presence of two-point correlations in all 
real systems (the classical SRO), which could obviously mask the speckle 
distribution analysis.
In this respect, we have checked that, at least for SRO lattices with no 
high-order correlations, 
dividing the speckles  pattern intensity by its associated diffuse scattering 
one $I_{DD}$ (by smoothing, averaging or fitting) makes $P(I/I_{DD})$ 
follow the normal decreasing exponential law.
This can help disentangling high-orders effects from two-point ones.

Another issue might be the partial coherence of the beam, which reduces the 
speckles contrast and makes the previous analyses difficult.
This could be overcome by the analysis of the speckles maximum intensities, which
exhibit similar statistical properties (not shown here).

In conclusion, we suggest the that speckle statistics analysis could be used 
as another tool to test for the presence of high-order correlations.
Indeed much physics could be explored with the measurement of high-order 
correlations, still hidden to experiments (see ref. \cite{Treacy} for examples).
We hope this work will impulse theoretical and experimental studies on the role
of high-order correlation functions on coherent diffraction diagram.

\begin{acknowledgments}
We thank F. Berenguer, D. Gratias, D. Le Bolloc'h and F. Livet for 
useful discussions.
\end{acknowledgments}

\bibliography{homometry}

\begin{thebibliography}{32}%
\makeatletter
\providecommand \@ifxundefined [1]{%
 \@ifx{#1\undefined}
}%
\providecommand \@ifnum [1]{%
 \ifnum #1\expandafter \@firstoftwo
 \else \expandafter \@secondoftwo
 \fi
}%
\providecommand \@ifx [1]{%
 \ifx #1\expandafter \@firstoftwo
 \else \expandafter \@secondoftwo
 \fi
}%
\providecommand \natexlab [1]{#1}%
\providecommand \enquote  [1]{``#1''}%
\providecommand \bibnamefont  [1]{#1}%
\providecommand \bibfnamefont [1]{#1}%
\providecommand \citenamefont [1]{#1}%
\providecommand \href@noop [0]{\@secondoftwo}%
\providecommand \href [0]{\begingroup \@sanitize@url \@href}%
\providecommand \@href[1]{\@@startlink{#1}\@@href}%
\providecommand \@@href[1]{\endgroup#1\@@endlink}%
\providecommand \@sanitize@url [0]{\catcode `\\12\catcode `\$12\catcode
  `\&12\catcode `\#12\catcode `\^12\catcode `\_12\catcode `\%12\relax}%
\providecommand \@@startlink[1]{}%
\providecommand \@@endlink[0]{}%
\providecommand \url  [0]{\begingroup\@sanitize@url \@url }%
\providecommand \@url [1]{\endgroup\@href {#1}{\urlprefix }}%
\providecommand \urlprefix  [0]{URL }%
\providecommand \Eprint [0]{\href }%
\providecommand \doibase [0]{http://dx.doi.org/}%
\providecommand \selectlanguage [0]{\@gobble}%
\providecommand \bibinfo  [0]{\@secondoftwo}%
\providecommand \bibfield  [0]{\@secondoftwo}%
\providecommand \translation [1]{[#1]}%
\providecommand \BibitemOpen [0]{}%
\providecommand \bibitemStop [0]{}%
\providecommand \bibitemNoStop [0]{.\EOS\space}%
\providecommand \EOS [0]{\spacefactor3000\relax}%
\providecommand \BibitemShut  [1]{\csname bibitem#1\endcsname}%
\let\auto@bib@innerbib\@empty
\bibitem [{\citenamefont {Sutton}\ \emph {et~al.}(1991)\citenamefont {Sutton},
  \citenamefont {Mochrie}, \citenamefont {Greytak}, \citenamefont {Nagler},
  \citenamefont {Berman}, \citenamefont {Held},\ and\ \citenamefont
  {Stephenson}}]{Sutton}%
  \BibitemOpen
  \bibfield  {author} {\bibinfo {author} {\bibfnamefont {M.}~\bibnamefont
  {Sutton}}, \bibinfo {author} {\bibfnamefont {S.~G.~J.}\ \bibnamefont
  {Mochrie}}, \bibinfo {author} {\bibfnamefont {T.}~\bibnamefont {Greytak}},
  \bibinfo {author} {\bibfnamefont {S.~E.}\ \bibnamefont {Nagler}}, \bibinfo
  {author} {\bibfnamefont {L.~E.}\ \bibnamefont {Berman}}, \bibinfo {author}
  {\bibfnamefont {G.~E.}\ \bibnamefont {Held}}, \ and\ \bibinfo {author}
  {\bibfnamefont {G.~B.}\ \bibnamefont {Stephenson}},\ }\href@noop {}
  {\bibfield  {journal} {\bibinfo  {journal} {Nature (London)}\ }\textbf
  {\bibinfo {volume} {352}},\ \bibinfo {pages} {608} (\bibinfo {year}
  {1991})}\BibitemShut {NoStop}%
\bibitem [{\citenamefont {Vartanyants}\ \emph {et~al.}(2011)\citenamefont
  {Vartanyants}, \citenamefont {Singer}, \citenamefont {Mancuso}, \citenamefont
  {Yefanov}, \citenamefont {Sakdinawat}, \citenamefont {Liu}, \citenamefont
  {Bang}, \citenamefont {Williams}, \citenamefont {Cadenazzi}, \citenamefont
  {Abbey}, \citenamefont {Sinn}, \citenamefont {Attwood}, \citenamefont
  {Nugent}, \citenamefont {Weckert}, \citenamefont {Wang}, \citenamefont {Zhu},
  \citenamefont {Wu}, \citenamefont {Graves}, \citenamefont {Scherz},
  \citenamefont {Turner}, \citenamefont {Schlotter}, \citenamefont
  {Messerschmidt}, \citenamefont {{L\"uning}}, \citenamefont {Acremann},
  \citenamefont {Heimann}, \citenamefont {Mancini}, \citenamefont {Joshi},
  \citenamefont {Krzywinski}, \citenamefont {Soufli}, \citenamefont
  {Fernandez-Perea}, \citenamefont {Hau-Riege}, \citenamefont {Peele},
  \citenamefont {Feng}, \citenamefont {Krupin}, \citenamefont {Moeller},\ and\
  \citenamefont {Wurth}}]{Vartanyants2011}%
  \BibitemOpen
  \bibfield  {author} {\bibinfo {author} {\bibfnamefont {I.~A.}\ \bibnamefont
  {Vartanyants}}, \bibinfo {author} {\bibfnamefont {A.}~\bibnamefont {Singer}},
  \bibinfo {author} {\bibfnamefont {A.}~\bibnamefont {Mancuso}}, \bibinfo
  {author} {\bibfnamefont {O.}~\bibnamefont {Yefanov}}, \bibinfo {author}
  {\bibfnamefont {A.}~\bibnamefont {Sakdinawat}}, \bibinfo {author}
  {\bibfnamefont {Y.}~\bibnamefont {Liu}}, \bibinfo {author} {\bibfnamefont
  {E.}~\bibnamefont {Bang}}, \bibinfo {author} {\bibfnamefont {G.}~\bibnamefont
  {Williams}}, \bibinfo {author} {\bibfnamefont {G.}~\bibnamefont {Cadenazzi}},
  \bibinfo {author} {\bibfnamefont {B.}~\bibnamefont {Abbey}}, \bibinfo
  {author} {\bibfnamefont {H.}~\bibnamefont {Sinn}}, \bibinfo {author}
  {\bibfnamefont {D.}~\bibnamefont {Attwood}}, \bibinfo {author} {\bibfnamefont
  {K.}~\bibnamefont {Nugent}}, \bibinfo {author} {\bibfnamefont
  {E.}~\bibnamefont {Weckert}}, \bibinfo {author} {\bibfnamefont
  {T.}~\bibnamefont {Wang}}, \bibinfo {author} {\bibfnamefont {D.}~\bibnamefont
  {Zhu}}, \bibinfo {author} {\bibfnamefont {B.}~\bibnamefont {Wu}}, \bibinfo
  {author} {\bibfnamefont {C.}~\bibnamefont {Graves}}, \bibinfo {author}
  {\bibfnamefont {A.}~\bibnamefont {Scherz}}, \bibinfo {author} {\bibfnamefont
  {J.}~\bibnamefont {Turner}}, \bibinfo {author} {\bibfnamefont
  {W.}~\bibnamefont {Schlotter}}, \bibinfo {author} {\bibfnamefont
  {M.}~\bibnamefont {Messerschmidt}}, \bibinfo {author} {\bibfnamefont
  {J.}~\bibnamefont {{L\"uning}}}, \bibinfo {author} {\bibfnamefont
  {Y.}~\bibnamefont {Acremann}}, \bibinfo {author} {\bibfnamefont
  {P.}~\bibnamefont {Heimann}}, \bibinfo {author} {\bibfnamefont
  {D.}~\bibnamefont {Mancini}}, \bibinfo {author} {\bibfnamefont
  {V.}~\bibnamefont {Joshi}}, \bibinfo {author} {\bibfnamefont
  {J.}~\bibnamefont {Krzywinski}}, \bibinfo {author} {\bibfnamefont
  {R.}~\bibnamefont {Soufli}}, \bibinfo {author} {\bibfnamefont
  {M.}~\bibnamefont {Fernandez-Perea}}, \bibinfo {author} {\bibfnamefont
  {S.}~\bibnamefont {Hau-Riege}}, \bibinfo {author} {\bibfnamefont
  {A.}~\bibnamefont {Peele}}, \bibinfo {author} {\bibfnamefont
  {Y.}~\bibnamefont {Feng}}, \bibinfo {author} {\bibfnamefont {O.}~\bibnamefont
  {Krupin}}, \bibinfo {author} {\bibfnamefont {W.}~\bibnamefont {Moeller}}, \
  and\ \bibinfo {author} {\bibfnamefont {W.}~\bibnamefont {Wurth}},\
  }\href@noop {} {\bibfield  {journal} {\bibinfo  {journal} {Phys. Rev. Lett.}\
  }\textbf {\bibinfo {volume} {107}},\ \bibinfo {pages} {144801} (\bibinfo
  {year} {2011})}\BibitemShut {NoStop}%
\bibitem [{\citenamefont {Livet}(2007)}]{Livet}%
  \BibitemOpen
  \bibfield  {author} {\bibinfo {author} {\bibfnamefont {F.}~\bibnamefont
  {Livet}},\ }\href@noop {} {\bibfield  {journal} {\bibinfo  {journal} {Acta
  Cryst. A}\ }\textbf {\bibinfo {volume} {63}},\ \bibinfo {pages} {87}
  (\bibinfo {year} {2007})}\BibitemShut {NoStop}%
\bibitem [{\citenamefont {Miao}\ \emph {et~al.}()\citenamefont {Miao},
  \citenamefont {Sayre},\ and\ \citenamefont {Chapman}}]{Miao}%
  \BibitemOpen
  \bibfield  {author} {\bibinfo {author} {\bibfnamefont {J.}~\bibnamefont
  {Miao}}, \bibinfo {author} {\bibfnamefont {D.}~\bibnamefont {Sayre}}, \ and\
  \bibinfo {author} {\bibfnamefont {H.}~\bibnamefont {Chapman}},\ }\href@noop
  {} {\bibfield  {journal} {\bibinfo  {journal} {J. Opt. Soc. Am. A}\ }\textbf
  {\bibinfo {volume} {15}}}\BibitemShut {NoStop}%
\bibitem [{\citenamefont {Rotenburg}\ and\ \citenamefont
  {Faulkner}(2004)}]{Rodenburg}%
  \BibitemOpen
  \bibfield  {author} {\bibinfo {author} {\bibfnamefont {J.~M.}\ \bibnamefont
  {Rotenburg}}\ and\ \bibinfo {author} {\bibfnamefont {H.~M.}\ \bibnamefont
  {Faulkner}},\ }\href@noop {} {\bibfield  {journal} {\bibinfo  {journal}
  {Appl. Phys. Lett.}\ }\textbf {\bibinfo {volume} {85}},\ \bibinfo {pages}
  {4795} (\bibinfo {year} {2004})}\BibitemShut {NoStop}%
\bibitem [{\citenamefont {Ravy}\ \emph {et~al.}()\citenamefont {Ravy},
  \citenamefont {Bolloc'h}, \citenamefont {Currat}, \citenamefont {Fluerasu},
  \citenamefont {Mocuta},\ and\ \citenamefont {Dkhil}}]{Ravy2007}%
  \BibitemOpen
  \bibfield  {author} {\bibinfo {author} {\bibfnamefont {S.}~\bibnamefont
  {Ravy}}, \bibinfo {author} {\bibfnamefont {D.~L.}\ \bibnamefont {Bolloc'h}},
  \bibinfo {author} {\bibfnamefont {R.}~\bibnamefont {Currat}}, \bibinfo
  {author} {\bibfnamefont {A.}~\bibnamefont {Fluerasu}}, \bibinfo {author}
  {\bibfnamefont {C.}~\bibnamefont {Mocuta}}, \ and\ \bibinfo {author}
  {\bibfnamefont {B.}~\bibnamefont {Dkhil}},\ }\href@noop {} {\bibfield
  {journal} {\bibinfo  {journal} {Phys. Rev. Lett.}\ }\textbf {\bibinfo
  {volume} {98}}}\BibitemShut {NoStop}%
\bibitem [{\citenamefont {{Gr\"ubel}}\ and\ \citenamefont
  {Zontone}(2004)}]{Grubel2004}%
  \BibitemOpen
  \bibfield  {author} {\bibinfo {author} {\bibfnamefont {G.}~\bibnamefont
  {{Gr\"ubel}}}\ and\ \bibinfo {author} {\bibfnamefont {F.}~\bibnamefont
  {Zontone}},\ }\href@noop {} {\bibfield  {journal} {\bibinfo  {journal} {J.
  Alloys Compd.}\ }\textbf {\bibinfo {volume} {362}},\ \bibinfo {pages} {3}
  (\bibinfo {year} {2004})}\BibitemShut {NoStop}%
\bibitem [{\citenamefont {Wochner}\ \emph {et~al.}(2009)\citenamefont
  {Wochner}, \citenamefont {Gutt}, \citenamefont {Autenrieth}, \citenamefont
  {Demmer}, \citenamefont {Bugaev}, \citenamefont {Ortiz}, \citenamefont
  {Duri}, \citenamefont {Zontone}, \citenamefont {{Gr\"ubel}},\ and\
  \citenamefont {Dosch}}]{Wochner2009}%
  \BibitemOpen
  \bibfield  {author} {\bibinfo {author} {\bibfnamefont {P.}~\bibnamefont
  {Wochner}}, \bibinfo {author} {\bibfnamefont {C.}~\bibnamefont {Gutt}},
  \bibinfo {author} {\bibfnamefont {T.}~\bibnamefont {Autenrieth}}, \bibinfo
  {author} {\bibfnamefont {T.}~\bibnamefont {Demmer}}, \bibinfo {author}
  {\bibfnamefont {V.}~\bibnamefont {Bugaev}}, \bibinfo {author} {\bibfnamefont
  {A.~D.}\ \bibnamefont {Ortiz}}, \bibinfo {author} {\bibfnamefont
  {A.}~\bibnamefont {Duri}}, \bibinfo {author} {\bibfnamefont {F.}~\bibnamefont
  {Zontone}}, \bibinfo {author} {\bibfnamefont {G.}~\bibnamefont {{Gr\"ubel}}},
  \ and\ \bibinfo {author} {\bibfnamefont {H.}~\bibnamefont {Dosch}},\
  }\href@noop {} {\bibfield  {journal} {\bibinfo  {journal} {Proc. Natl. Acad.
  Sci. U.S.A.}\ }\textbf {\bibinfo {volume} {106}},\ \bibinfo {pages} {11511}
  (\bibinfo {year} {2009})}\BibitemShut {NoStop}%
\bibitem [{\citenamefont {Altarelli}\ \emph {et~al.}(2010)\citenamefont
  {Altarelli}, \citenamefont {Kurta},\ and\ \citenamefont
  {Vartanyants}}]{Altarelli2010}%
  \BibitemOpen
  \bibfield  {author} {\bibinfo {author} {\bibfnamefont {M.}~\bibnamefont
  {Altarelli}}, \bibinfo {author} {\bibfnamefont {R.~P.}\ \bibnamefont
  {Kurta}}, \ and\ \bibinfo {author} {\bibfnamefont {I.~A.}\ \bibnamefont
  {Vartanyants}},\ }\href@noop {} {\bibfield  {journal} {\bibinfo  {journal}
  {Phys. Rev. B}\ }\textbf {\bibinfo {volume} {82}},\ \bibinfo {pages} {104207}
  (\bibinfo {year} {2010})}\BibitemShut {NoStop}%
\bibitem [{\citenamefont {Su}\ \emph {et~al.}(2011)\citenamefont {Su},
  \citenamefont {Seu}, \citenamefont {Parks}, \citenamefont {Kan},
  \citenamefont {Fullerton}, \citenamefont {Roy},\ and\ \citenamefont
  {Kevan}}]{Su2011}%
  \BibitemOpen
  \bibfield  {author} {\bibinfo {author} {\bibfnamefont {R.}~\bibnamefont
  {Su}}, \bibinfo {author} {\bibfnamefont {K.~A.}\ \bibnamefont {Seu}},
  \bibinfo {author} {\bibfnamefont {D.}~\bibnamefont {Parks}}, \bibinfo
  {author} {\bibfnamefont {J.~J.}\ \bibnamefont {Kan}}, \bibinfo {author}
  {\bibfnamefont {E.~E.}\ \bibnamefont {Fullerton}}, \bibinfo {author}
  {\bibfnamefont {S.}~\bibnamefont {Roy}}, \ and\ \bibinfo {author}
  {\bibfnamefont {S.~D.}\ \bibnamefont {Kevan}},\ }\href@noop {} {\bibfield
  {journal} {\bibinfo  {journal} {Phys. Rev. Lett.}\ }\textbf {\bibinfo
  {volume} {107}} (\bibinfo {year} {2011})}\BibitemShut {NoStop}%
\bibitem [{\citenamefont {Axel}\ \emph {et~al.}(1992)\citenamefont {Axel},
  \citenamefont {Allouche},\ and\ \citenamefont {Went}}]{Axel1992}%
  \BibitemOpen
  \bibfield  {author} {\bibinfo {author} {\bibfnamefont {F.}~\bibnamefont
  {Axel}}, \bibinfo {author} {\bibfnamefont {J.}~\bibnamefont {Allouche}}, \
  and\ \bibinfo {author} {\bibfnamefont {Z.}~\bibnamefont {Went}},\ }\href@noop
  {} {\bibfield  {journal} {\bibinfo  {journal} {J. Phys.: Condens. Matter}\
  }\textbf {\bibinfo {volume} {4}},\ \bibinfo {pages} {8713} (\bibinfo {year}
  {1992})}\BibitemShut {NoStop}%
\bibitem [{\citenamefont {Baake}\ and\ \citenamefont
  {Grimm}(2009)}]{Baake2009}%
  \BibitemOpen
  \bibfield  {author} {\bibinfo {author} {\bibfnamefont {M.}~\bibnamefont
  {Baake}}\ and\ \bibinfo {author} {\bibfnamefont {U.}~\bibnamefont {Grimm}},\
  }\href@noop {} {\bibfield  {journal} {\bibinfo  {journal} {Phys. Rev. B}\
  }\textbf {\bibinfo {volume} {79}},\ \bibinfo {pages} {020203(R)} (\bibinfo
  {year} {2009})}\BibitemShut {NoStop}%
\bibitem [{\citenamefont {Guinier}(1994)}]{Guinier94}%
  \BibitemOpen
  \bibfield  {author} {\bibinfo {author} {\bibfnamefont {A.}~\bibnamefont
  {Guinier}},\ }\href@noop {} {\emph {\bibinfo {title} {X-Ray Diffraction in
  Crystals, Imperfect Crystals, and Amorphous Bodies.}}}\ (\bibinfo
  {publisher} {Dover Publications},\ \bibinfo {address} {New York},\ \bibinfo
  {year} {1994})\BibitemShut {NoStop}%
\bibitem [{\citenamefont {Welberry}(1994)}]{Welberry1994}%
  \BibitemOpen
  \bibfield  {author} {\bibinfo {author} {\bibfnamefont {T.~R.}\ \bibnamefont
  {Welberry}},\ }\href@noop {} {\bibfield  {journal} {\bibinfo  {journal} {J.
  Appl. Cryst.}\ }\textbf {\bibinfo {volume} {27}},\ \bibinfo {pages} {205}
  (\bibinfo {year} {1994})}\BibitemShut {NoStop}%
\bibitem [{\citenamefont {Patterson}(1939)}]{Patterson}%
  \BibitemOpen
  \bibfield  {author} {\bibinfo {author} {\bibfnamefont {A.~L.}\ \bibnamefont
  {Patterson}},\ }\href@noop {} {\bibfield  {journal} {\bibinfo  {journal}
  {Nature (London)}\ }\textbf {\bibinfo {volume} {143}},\ \bibinfo {pages}
  {939} (\bibinfo {year} {1939})}\BibitemShut {NoStop}%
\bibitem [{\citenamefont {Patterson}(1944)}]{Patterson44}%
  \BibitemOpen
  \bibfield  {author} {\bibinfo {author} {\bibfnamefont {A.~L.}\ \bibnamefont
  {Patterson}},\ }\href@noop {} {\bibfield  {journal} {\bibinfo  {journal}
  {Phys. Rev.}\ }\textbf {\bibinfo {volume} {65}},\ \bibinfo {pages} {195}
  (\bibinfo {year} {1944})}\BibitemShut {NoStop}%
\bibitem [{\citenamefont {Senechal}(2008)}]{Senechal}%
  \BibitemOpen
  \bibfield  {author} {\bibinfo {author} {\bibfnamefont {M.}~\bibnamefont
  {Senechal}},\ }\href@noop {} {\bibfield  {journal} {\bibinfo  {journal}
  {European Journal of Combinatorics}\ }\textbf {\bibinfo {volume} {29}},\
  \bibinfo {pages} {1933} (\bibinfo {year} {2008})}\BibitemShut {NoStop}%
\bibitem [{\citenamefont {Sayre}(1952)}]{Sayre}%
  \BibitemOpen
  \bibfield  {author} {\bibinfo {author} {\bibfnamefont {D.}~\bibnamefont
  {Sayre}},\ }\href@noop {} {\bibfield  {journal} {\bibinfo  {journal} {Acta
  Cryst. A}\ }\textbf {\bibinfo {volume} {5}},\ \bibinfo {pages} {843}
  (\bibinfo {year} {1952})}\BibitemShut {NoStop}%
\bibitem [{\citenamefont {van~der Veen}\ and\ \citenamefont
  {Pfeiffer}(2004)}]{vanderVeen}%
  \BibitemOpen
  \bibfield  {author} {\bibinfo {author} {\bibfnamefont {F.}~\bibnamefont
  {van~der Veen}}\ and\ \bibinfo {author} {\bibfnamefont {F.}~\bibnamefont
  {Pfeiffer}},\ }\href@noop {} {\bibfield  {journal} {\bibinfo  {journal} {J.
  Phys.: Condens. Matter}\ }\textbf {\bibinfo {volume} {16}},\ \bibinfo {pages}
  {5003} (\bibinfo {year} {2004})}\BibitemShut {NoStop}%
\bibitem [{\citenamefont {Welberry}(1977)}]{Welberry1977}%
  \BibitemOpen
  \bibfield  {author} {\bibinfo {author} {\bibfnamefont {T.~R.}\ \bibnamefont
  {Welberry}},\ }\href@noop {} {\bibfield  {journal} {\bibinfo  {journal} {J.
  Appl. Cryst.}\ }\textbf {\bibinfo {volume} {10}},\ \bibinfo {pages} {344}
  (\bibinfo {year} {1977})}\BibitemShut {NoStop}%
\bibitem [{Note1()}]{Note1}%
  \BibitemOpen
  \bibinfo {note} {In the following, we consider that an order parameter is
  short-range ordered (SRO) if its associated CF $g(n)$ vanishes at infinity
  and long-range ordered (LRO) otherwise.}\BibitemShut {Stop}%
\bibitem [{\citenamefont {Gratias}\ \emph {et~al.}(2005)\citenamefont
  {Gratias}, \citenamefont {Bresson},\ and\ \citenamefont
  {Quiquandon}}]{Gratias2005}%
  \BibitemOpen
  \bibfield  {author} {\bibinfo {author} {\bibfnamefont {D.}~\bibnamefont
  {Gratias}}, \bibinfo {author} {\bibfnamefont {L.}~\bibnamefont {Bresson}}, \
  and\ \bibinfo {author} {\bibfnamefont {M.}~\bibnamefont {Quiquandon}},\
  }\href@noop {} {\bibfield  {journal} {\bibinfo  {journal} {Annu. Rev. Mater.
  Res.}\ }\textbf {\bibinfo {volume} {35}},\ \bibinfo {pages} {75} (\bibinfo
  {year} {2005})}\BibitemShut {NoStop}%
\bibitem [{\citenamefont {Gratias}\ and\ \citenamefont
  {Axel}(1995)}]{beyond95}%
  \BibitemOpen
  \bibinfo {editor} {\bibfnamefont {D.}~\bibnamefont {Gratias}}\ and\ \bibinfo
  {editor} {\bibfnamefont {F.}~\bibnamefont {Axel}},\ eds.,\ \href@noop {}
  {\emph {\bibinfo {title} {Beyond quasicrystals.}}}\ (\bibinfo  {publisher}
  {Springer Verlag},\ \bibinfo {address} {Berlin},\ \bibinfo {year}
  {1995})\BibitemShut {NoStop}%
\bibitem [{\citenamefont {{H\"offe}}\ and\ \citenamefont
  {Baake}(2000)}]{Hoffe}%
  \BibitemOpen
  \bibfield  {author} {\bibinfo {author} {\bibfnamefont {M.}~\bibnamefont
  {{H\"offe}}}\ and\ \bibinfo {author} {\bibfnamefont {M.}~\bibnamefont
  {Baake}},\ }\href@noop {} {\bibfield  {journal} {\bibinfo  {journal} {Z.
  Krystallogr.}\ }\textbf {\bibinfo {volume} {215}},\ \bibinfo {pages} {441}
  (\bibinfo {year} {2000})}\BibitemShut {NoStop}%
\bibitem [{\citenamefont {Dainty}(1976)}]{Dainty1976}%
  \BibitemOpen
  \bibfield  {author} {\bibinfo {author} {\bibfnamefont {J.~C.}\ \bibnamefont
  {Dainty}},\ }\enquote {\bibinfo {title} {Progress in optics, vol xiv},}\ \
  (\bibinfo  {publisher} {North Holland},\ \bibinfo {address} {Amsterdam},\
  \bibinfo {year} {1976})\ Chap.\ \bibinfo {chapter} {The statistics of speckle
  patterns}\BibitemShut {NoStop}%
\bibitem [{Note2()}]{Note2}%
  \BibitemOpen
  \bibinfo {note} {We checked that RS triplet CF $g_3(n,n')=\protect \overline
  {\sigma _0\sigma _n\sigma _{n'}} \sim 0$ and do not show any structure, so
  that quadruplet are the first relevant RS high-order CF. Amongst them, we
  chose $g_4(n)$ because it is reminiscent of the very definition of $\sigma
  _n$.}\BibitemShut {Stop}%
\bibitem [{Note3()}]{Note3}%
  \BibitemOpen
  \bibinfo {note} {This behavior has been checked up to the $N=2^{14}$ RS
  sequences}\BibitemShut {NoStop}%
\bibitem [{\citenamefont {Baake}\ and\ \citenamefont
  {Grimm}(2011)}]{Baake2010}%
  \BibitemOpen
  \bibfield  {author} {\bibinfo {author} {\bibfnamefont {M.}~\bibnamefont
  {Baake}}\ and\ \bibinfo {author} {\bibfnamefont {U.}~\bibnamefont {Grimm}},\
  }\href@noop {} {\bibfield  {journal} {\bibinfo  {journal} {Phil. Mag.}\
  }\textbf {\bibinfo {volume} {61}},\ \bibinfo {pages} {2661} (\bibinfo {year}
  {2011})}\BibitemShut {NoStop}%
\bibitem [{\citenamefont {Guizar-Sicairos}\ \emph {et~al.}(2010)\citenamefont
  {Guizar-Sicairos}, \citenamefont {Evans-Lutterodt}, \citenamefont {Isakovic},
  \citenamefont {Stein}, \citenamefont {Warren}, \citenamefont {Sandy},
  \citenamefont {Narayanan},\ and\ \citenamefont {Fienup}}]{Guizar}%
  \BibitemOpen
  \bibfield  {author} {\bibinfo {author} {\bibfnamefont {M.}~\bibnamefont
  {Guizar-Sicairos}}, \bibinfo {author} {\bibfnamefont {K.}~\bibnamefont
  {Evans-Lutterodt}}, \bibinfo {author} {\bibfnamefont {A.~F.}\ \bibnamefont
  {Isakovic}}, \bibinfo {author} {\bibfnamefont {A.}~\bibnamefont {Stein}},
  \bibinfo {author} {\bibfnamefont {J.~B.}\ \bibnamefont {Warren}}, \bibinfo
  {author} {\bibfnamefont {A.~R.}\ \bibnamefont {Sandy}}, \bibinfo {author}
  {\bibfnamefont {S.}~\bibnamefont {Narayanan}}, \ and\ \bibinfo {author}
  {\bibfnamefont {J.~R.}\ \bibnamefont {Fienup}},\ }\href@noop {} {\bibfield
  {journal} {\bibinfo  {journal} {Optics Express}\ }\textbf {\bibinfo {volume}
  {18}},\ \bibinfo {pages} {18374} (\bibinfo {year} {2010})}\BibitemShut
  {NoStop}%
\bibitem [{\citenamefont {Schneider}\ \emph {et~al.}(2010)\citenamefont
  {Schneider}, \citenamefont {Seibald}, \citenamefont {Lagally},\ and\
  \citenamefont {Oeckler}}]{Schneider}%
  \BibitemOpen
  \bibfield  {author} {\bibinfo {author} {\bibfnamefont {M.}~\bibnamefont
  {Schneider}}, \bibinfo {author} {\bibfnamefont {M.}~\bibnamefont {Seibald}},
  \bibinfo {author} {\bibfnamefont {P.}~\bibnamefont {Lagally}}, \ and\
  \bibinfo {author} {\bibfnamefont {O.}~\bibnamefont {Oeckler}},\ }\href@noop
  {} {\bibfield  {journal} {\bibinfo  {journal} {J. Appl. Cryst.}\ }\textbf
  {\bibinfo {volume} {43}},\ \bibinfo {pages} {1012} (\bibinfo {year}
  {2010})}\BibitemShut {NoStop}%
\bibitem [{\citenamefont {Treacy}\ \emph {et~al.}(2005)\citenamefont {Treacy},
  \citenamefont {Gibson}, \citenamefont {Fan}, \citenamefont {Patterson},\ and\
  \citenamefont {McNulty}}]{Treacy}%
  \BibitemOpen
  \bibfield  {author} {\bibinfo {author} {\bibfnamefont {M.~M.~J.}\
  \bibnamefont {Treacy}}, \bibinfo {author} {\bibfnamefont {J.}~\bibnamefont
  {Gibson}}, \bibinfo {author} {\bibfnamefont {L.}~\bibnamefont {Fan}},
  \bibinfo {author} {\bibfnamefont {D.~J.}\ \bibnamefont {Patterson}}, \ and\
  \bibinfo {author} {\bibfnamefont {I.}~\bibnamefont {McNulty}},\ }\href@noop
  {} {\bibfield  {journal} {\bibinfo  {journal} {Rep. Prog. Phys.}\ }\textbf
  {\bibinfo {volume} {68}},\ \bibinfo {pages} {2899} (\bibinfo {year}
  {2005})}\BibitemShut {NoStop}%
\bibitem [{Note4()}]{Note4}%
  \BibitemOpen
  \bibinfo {note} {Let us emphasize that the techniques developed in ref. \cite
  {Treacy} to get high-order CF require sample scanning, as in
  ptychography.}\BibitemShut {Stop}%
\end{thebibliography}%

\end{document}